\documentclass{article}
\usepackage[T1]{fontenc} 
\usepackage[utf8]{inputenc} 
\usepackage{ismir,amsmath,cite,url}
\usepackage{graphicx}
\usepackage{color}
\usepackage{booktabs}
\usepackage{amsfonts}

\usepackage{lineno}

\title{Learning Hierarchical Metrical Structure Beyond Measures}

\multauthor
{Junyan Jiang$^{1, 2}$ \hspace{1cm} Daniel Chin$^{1, 2}$ \hspace{1cm} Yixiao Zhang$^{1, 3}$ \hspace{1cm} Gus Xia$^{1, 2}$}{
$^1$ Music X Lab, NYU Shanghai \hspace{1cm}$^2$ MBZUAI \hspace{1cm} $^3$ Centre for Digital Music, QMUL\\
{\tt\small jj2731@nyu.edu, daniel.chin@nyu.edu, yixiao.zhang@qmul.ac.uk, gxia@nyu.edu}
}

\def\authorname{J. Jiang, D. Chin, Y. Zhang and G. Xia}

\usepackage[bookmarks=false,pdfauthor={\authorname},pdfsubject={\papersubject},hidelinks]{hyperref}

\begin{document}

\maketitle
\begin{abstract}
Music contains hierarchical structures beyond beats and measures. While hierarchical structure annotations are helpful for music information retrieval and computer musicology, such annotations are scarce in current digital music databases. In this paper, we explore a data-driven approach to automatically extract hierarchical metrical structures from scores. We propose a new model with a Temporal Convolutional Network-Conditional Random Field (TCN-CRF) architecture. Given a symbolic music score, our model takes in an arbitrary number of voices in a beat-quantized form, and predicts a 4-level hierarchical metrical structure from downbeat-level to section-level. We also annotate a dataset using RWC-POP MIDI files to facilitate training and evaluation. We show by experiments that the proposed method performs better than the rule-based approach under different orchestration settings. We also perform some simple musicological analysis on the model predictions. All demos, datasets and pre-trained models are publicly available on Github\footnote{\href{https://github.com/music-x-lab/Hierarchical-Metrical-Structure}{https://github.com/music-x-lab/Hierarchical-Metrical-Structure}}.
\end{abstract}
\section{Introduction}\label{sec:introduction}

Music contains rich structures at different levels, and the structure annotations play an important role in music understanding \cite{maddage2004content, ellis2007identifyingcover} and generation \cite{dieleman2018challenge, lattner2018imposing}. Progress has been made in music structure analysis on certain levels, like beat tracking \cite{ellis2007beat, beatroot, matthewdavies2019temporal}, downbeat detection \cite{ krebs2015inferring,durand2016robust,fuentes2019music,gouyon2005review} and part segmentation \cite{grill2015music,mcfee2014analyzing,salamon2021deep,mccallum2019unsupervised}. These levels of structures are often inter-connected with each other, and can be described in a hierarchical way. For example, a part may contain several sections, each containing several measures. Measures can be further decomposed into beats.

Several views of hierarchical music structures are formally discussed in the Generative Theory of Tonal Music (GTTM) \cite{gttm}, including the grouping structure, the metrical structure, the time-span tree, and the prolongational tree, each focusing on different music properties (i.e., the grouping structure focuses more on melodic grouping while the metrical structure focuses on rhythmic patterns). Among these structures, we choose the metrical structure as the topic for two reasons: (1) For polyphonic music, the metrical structures of different voices are usually compatible, building up a common song-level metrical structure. This property makes data annotation easier and provides opportunities for self-supervision; (2) Some low-level metrical structures (e.g., beats and downbeats) are already annotated in music scores and most MIDI datasets, which can be a helpful source of supervision. In this paper, we will focus our analysis on pop songs, which often contain a well-defined hierarchy of metrical structures\cite{robins2017phrase}.

\begin{figure}
    \centering
    \includegraphics[width=\linewidth,clip, trim=0.0cm 7.4cm 0.0cm 7.4cm]{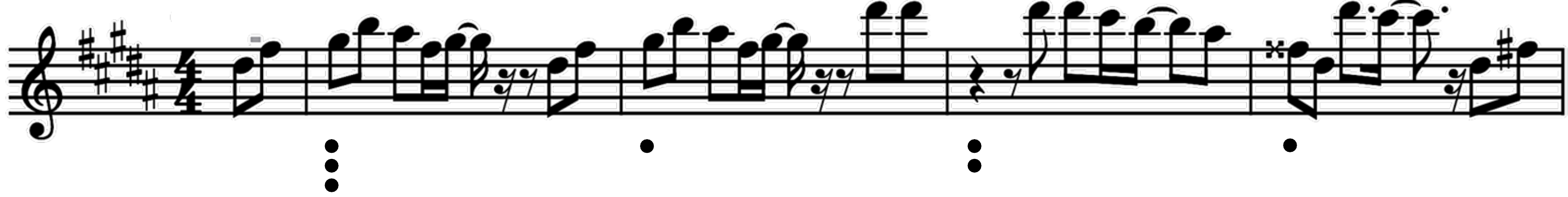}
    \caption{The hierarchical metrical structure of the beginning of RWC-POP No. 001. Only the main melody is shown. Despite the pick-up bar, these 4 measures are grouped into 2 hypermeasures of length 2, which are further grouped into 1 hypermeasure of length 4.}
    \label{fig:example}
\end{figure}

Our main goal is to infer the high-level metrical structures given low-level ones like beats and downbeats. The hierarchy of the metrical structure is created by recursively grouping lower metrical units into upper ones (see Figure \ref{fig:example} for an example). We call the grouping of measures (or other larger metrical units)  a {\it hypermeasure}, and the number of units that form the group as its {\it hypermeter} \cite{wojcik2010representations, temperley2008hypermetrical}. While some properties of beats and measures can be generalized to upper-level metrical structures, there are still many differences to take into account. Similar to the meter, the hypermeter of each layer tends to stay the same to maintain a regular rhythmic pulse, but it is not uncommon to see hypermeter changes in a piece, as shown in Figure \ref{fig:case}. Such changes occur more often than low-level meter changes since listeners are less sensitive to long-term rhythmic regularity. Another major difference is that the decision of upper-level metrical structures requires a longer context in the time domain compared to downbeat or beat tracking.

To resolve these issues, we design a new model that contains a Temporal Convolutional Network (TCN) front-end for metrical level prediction and a Conditional Random Field (CRF) decoder for joint metrical structure decoding. We design the transition of CRF hidden states to allow hypermeter changes with some penalties. To handle polyphony, the model takes an arbitrary number of voices/tracks as input, and predicts a confidence score for each track. The final prediction is a weighted average of the results from all tracks.  We annotated 70 songs from the RWC-POP dataset and used them for model training and evaluation. We conduct experiments under different orchestration setups with results shown in section \ref{sec:results}. 

\section{Related work}\label{sec:page_size}

\subsection{Downbeat and Meter Tracking}

Downbeat tracking for raw audio is a well-studied task with many promising results. 
Krebs et al. \cite{krebs2015inferring} use particle filters to find the downbeats, yielding a 2-level metrical structure. 
Durand et al. \cite{durand2016robust} use an ensemble of convolutional networks to locate downbeats robustly. 
Fuentes et al. \cite{fuentes2019music} use skip-chain CRF and deep learning to track downbeats, noting that longer-term musical contexts can better inform downbeat tracking. 
The reader is referred to Gouyon and Dixon's review \cite{gouyon2005review} for non-symbolic low-level metrical analyzers before 2005. 

Notably, locating the beats and downbeats for symbolic music is not a trivial task. 
Kostek et al. \cite{kostek2007searching} apply neural networks and rough sets to a polyphonic symbolic piece and classify whether each note is accented or not, yielding a 2-level metrical structure (beat and downbeat). 
Chuang and Su \cite{chuang2020beat} use various RNNs to classify each timestep in a piano roll into non-beat, beat, or downbeat. 

Another related task is time signature detection.
Benoit \cite{meudic2002automatic} uses symbolic-level auto-correlation to obtain a 4-level metrical structure, and ultimately extracts the meter. More auto-correlative methods \cite{brown1993determination, eck2005finding} share a similar logic, since note onsets usually display periodicity in every measure. 
More recently, inner metric analysis was also used to infer the time signature by Haas et al. \cite{de2016meter}. 

\subsection{Music Segmentation}
The task of music segmentation is to infer musically meaningful section or part boundaries from the music content. Audio-based music segmentation is usually achieved by detecting similarity or repetition of the audio spectral features.
McFee and Ellis \cite{mcfee2014analyzing} evaluate inter-time frame similarity and use spectral clustering to obtain a multi-level segmentation of music. 
Salamon et al. \cite{salamon2021deep} and McCallum \cite{mccallum2019unsupervised} replace traditional features with pre-trained deep embeddings to estimate timbre and harmonic similarity. Tralie and McFee \cite{tralie2019enhanced} fuse multiple similarity metrics for better prediction results. 
Ullrich et al. \cite{ullrich2014boundary} train fully supervised CNN on a segment-annotated dataset to detect segment boundaries. 
See Dannenberg and Goto's review \cite{dannenberg2008music} and the 2010 SOTA report by Paulus et al. \cite{paulus2010state} to learn more about audio-based music segmentations. 

Segmentation of symbolic music relies more on domain knowledge of music composition. 
Van der Werf and Hendriks \cite{van2004constraint} restate GTTM grouping rules in terms of Optimality Theory (OT) and design a Prolog program to find an optimal parse for short monophonic pieces. 
Dai et al. \cite{dai2020automatic} identify phrases as units of melodic repetition by minimizing the Structural Description Length (SDL) for the entire piece, and then extract a 2-layer hierarchy. 

\subsection{Hierarchical Structure Analysis}

The Generative Theory of Tonal Music (GTTM) \cite{gttm} discusses several views of the hierarchical music structures, but GTTM does not describe how to realize an analyzer computationally. Various efforts have been made to mechanize GTTM.
For example, Jones et al. \cite{jones1988rule} use a rule-based expert system to obtain a 6-level metrical structure for monophonic pieces, satisfying all GTTM's well-formedness rules while following Povel's grid theory \cite{povel1984theoretical}. 
Rosenthal's Machine Rhythm \cite{rosenthal1992machine} ventures into the polyphonic domain and uses rule-based methods to extract a 3-level rhythmic annotation for MIDI input. Temperley and Sleator \cite{temperley1999modeling} use a preference-rule approach and dynamic programming to obtain the optimal parse. Hamanaka et al. \cite{hamanaka2006implementing} propose the Automatic Time-span Tree Analyzer (ATTA) for structural analysis of 8-bar monophonic scores. Temperley \cite{temperley2009unified} use Bayesian reasoning to jointly analyze metrical, harmonic, and stream structures. Wojcik and Kostek \cite{wojcik2010representations} use rule-based methods to retrieve hypermetric rhythm from only the melody.

Machine learning models are also used for hierarchical structure analysis. Hamanaka et al. propose DeepGTTM \cite{hamanaka2016deepgttm, hiratadeepgttm} which is a fully automatic analyzer that uses a neural network to predict the applicability of GTTM rules for each note, yielding a 5-level metrical structure for 8-bar monophonic scores. 

There are some issues when applying previous works to large polyphonic MIDI databases. Most systems work only for monophonic music or music with limited polyphony. Also, they often work in a short context, usually up to 8 bars.

\section{Methodology}

\subsection{Problem Setting}

We first formally define the metrical structure prediction task for this paper. Assume we have a music score with $T$ voices (or tracks, in the sense of MIDI files) $\mathbf{m}_1, ..., \mathbf{m}_T$. We also have a list of pre-annotated downbeats $d_1, ..., d_N$ for the music. The aim is to assign hierarchical metrical labels $l_i \in \{0, 1, ..., L\}$ to each downbeat $d_i$ where $L$ is the total number of layers we want to build beyond measures. Each label serves as the level of the metrical boundary at $d_i$. $l_i=l$ means $d_i$ serves as a metrical boundary of all levels for the first $l$ levels beyond measures. Specially, $l_i=0$ means $d_i$ serves only as a measure boundary but not any metrical boundary beyond measures.

If we use GTTM's metrical structure notation, we can use $(l_i + 1)$ dots to represent a metrical boundary level of $l_i$. For the example in figure \ref{fig:example}, the first 4 measures (excluding the pickup measure) would have labels $l_1=2, l_2=0, l_3=1, l_4=0$.

We can also introduce the following notations. \\
\textbf{Hypermeasures}: A hypermeasure of level $l$ is an interval between two downbeats $d_i$ and $d_j$ where $l_i\geq l, l_j\geq l$ and $l_k<l$ for all $k=i+1...j-1$. In other words, any $l_k \geq l$ serves as a separator of a level-$l$ hypermeasure.  Specially, level-0 hypermeasures are just measures. \\
\textbf{Hypermeters}: A hypermeter is the generalization of meters by counting how many level-$(l - 1)$ hypermeasures are in a level-$l$ hypermeasure. Since a binary structure is the most commonly used \cite{temperley2008hypermetrical,rohrmeier2020towards}, we assume a general hypermeter of 2 in all levels, with a few exceptions that cause \textit{binary irregularity}. 


\begin{figure}[t]
    \centering
    \includegraphics[width=\linewidth,clip, trim=9.0cm 6.7cm 12.2cm 5.15cm, page=6]{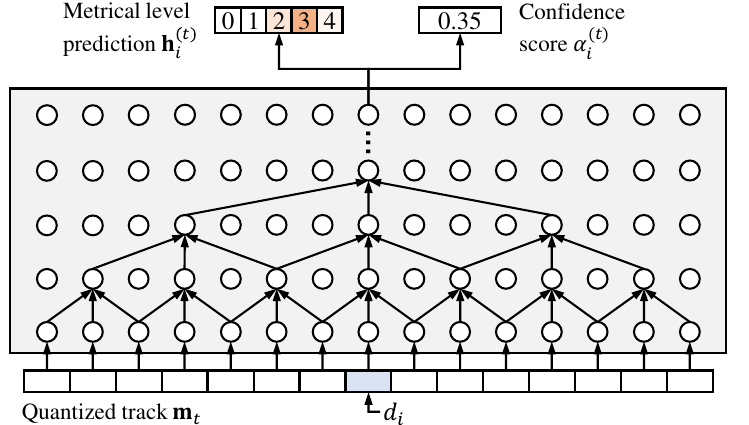}
    \caption{The architecture of the neural network.}
    \label{fig:tcn}
\end{figure}

\subsection{Temporal Convolutional Network}

Temporal Convolutional Networks (TCNs) have been proven an effective model for beat, downbeat, and tempo tracking \cite{matthewdavies2019temporal, bock2019multi, bock2020deconstruct}. We believe it is useful for general metrical structure analysis for its unique property we will mention below. We mainly reference \cite{bai2018empirical} for the design of the TCN, but we made it non-causal similar to \cite{matthewdavies2019temporal}. Each TCN block contains 8 sequential layers. The first 4 layers are a dilated convolutional layer with kernel size 3 and 256 channels, a batch normalization layer, a Rectified Linear Unit (ReLU) activation layer, and a dropout layer. The next 4 layers repeat this configuration. There is also a residual component that adds a linear transformation of the input to the block output, allowing shortcut connections.

Our model uses 6 TCN blocks sequentially. Each block multiplies the dilation by 2, starting from 1 at block 1, resulting in an exponentially growing context range for each layer. This allows the model to capture long-term context and more importantly, integrate prior knowledge about binary metrical structure into the network. The model input contains the piano roll and the onset roll of a track quantized into a 16th note level. Under a 4/4 meter, the dilations of the convolutional layers in each block are therefore 1/4 beat, 1/2 beat, 1 beat, 2 beats, 1 measure and 2 measures, respectively. This encourages the convolutional layers to capture more musically meaningful context for binary metrical structures.

For each track $\mathbf{m}_t$ in a song, we first feed them into the TCN blocks to get the features for each time step, and then discard the time steps that do not correspond to any downbeat. We use linear layers to project the features into a metrical level prediction $\mathbf{h}^{(t)}_i$ and a confidence score $\alpha^{(t)}_i$. Each prediction $\mathbf{h}^{(t)}_i$ is a vector of size $(L+1)$ for the labels $0...L$. The final prediction of $l_i$ is the weighted average of all predictions $\mathbf{h}^{(t)}_i$ on the same time step weighted by their confidence scores:
\begin{equation} \small
    a^{(t)}_i:=\exp{ \alpha^{(t)}_i} \big/ \sum_{t'}  { \exp{\alpha^{(t')}_i}}
\end{equation}
\vspace{-0.2 cm}
\begin{equation} \small
    \mathbf{p}_i:=\sum_t a^{(t)}_i \textrm{Softmax}(\mathbf{h}^{(t)}_i)
\end{equation}

\subsection{Conditional Random Fields}

Conditional Random Fields (CRFs) have been widely used in downbeat tracking \cite{durand2016downbeat, fuentes2019music} to enforce the regularity of the decoded downbeat patterns. Inspired by this, we also use a structured prediction method for decoding. The major differences are that this model needs to predict a hierarchy of $L=4$ layers of metrical structures jointly.

\begin{figure}[t]
    \centering
    \includegraphics[width=\linewidth,clip, trim=6.0cm 5.6cm 10.0cm 3.5cm]{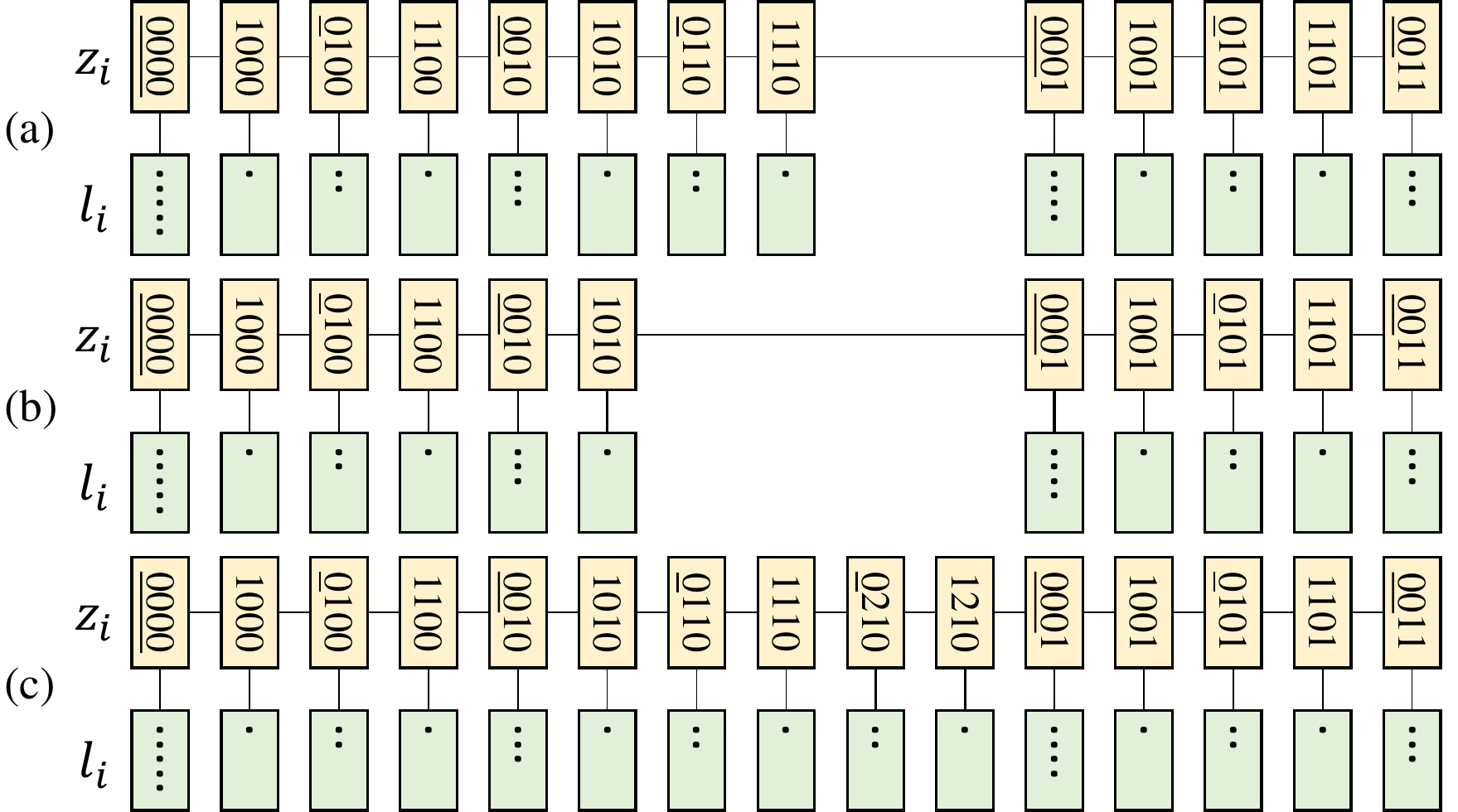}
    \caption{Examples of the CRF hidden variables $z_i$ (shown as a $L$-digit number $z_i^{(1)}...z_i^{(L)}$) and the corresponding metrical boundary levels $l_i$ when $L=4$. (a) Binary regularity is satisfied. (b) A level-$1$ hypermeasure is deleted from (a). (c) A level-$1$ hypermeasure is inserted into (a). Both (b) and (c) are examples of binary irregularity.}
    \label{fig:crf}
\end{figure}

The state space of hierarchical metrical structure can be complex and ambiguous. To make it simple, we restrict our model to accept a hypermeter of 1, 2 or 3 on any level. In the sense of transformational grammar, a level-$l$ hypermeter of 1 can be constructed by deleting some level-$(l-1)$ hypermeasures from a deep structure with binary regularity, and a hypermeter of 3 can be constructed by inserting some level-$(l-1)$ hypermeasures (see figure \ref{fig:crf} for an example). A hypermeter of 4 or more is not allowed and needs to be decomposed into multiple metrical levels (e.g., $4=2+2$).


We design the linear CRF with a joint state space $z_i=(z_i^{(1)}, ..., z_i^{(L)})$ where each $z_i^{(l)}\in\{0, 1, 2\}$ corresponds to the current hypermeasure position at level $l$, i.e., the number of complete level-$(l-1)$ hypermeasures in this level-$l$ hypermeasure up to the current time step. It can be seen as a generalization of the beat position in \cite{fuentes2019music}. A state $z_i^{(1...l)} = 0 \wedge z_i^{(l + 1)} \neq 0$ denotes a metrical boundary level $l_i = l$. Specially, the highest-level metrical boundary 
$l_i = L$ is associated and only associated with the state $(0, ..., 0)$, and the lowest-level metrical boundary $l_i = 0$ is associated with any $z_i$ where $z_i^{(1)} \neq 0$. See figure \ref{fig:crf} for some concrete examples.

We manually design the transition potential function to encode the belief of binary regularity on each level. We define the transition potential matrix of a single level as
\begin{equation} \small
    \mathbf{A}^{(l)} = \begin{bmatrix}
\exp{\big(-w_\mathrm{del}^{(l)}\big)} & 1 & 0\\
1 & 0 & \exp{\big(-w_\mathrm{ins}^{(l)}\big)}\\
1 & 0 & 0
\end{bmatrix}
\end{equation}
where $A^{(l)}_{ij}$ denotes the potential of transition from hypermeasure position $i$ to position $j$ on level $l$. $w_\mathrm{del}^{(l)} > 0, w_\mathrm{ins}^{(l)} > 0$ are hyperparameters that controls the penalty of a level-$l$ hypermeasure deletion and insertion respectively. Intuitively, an alternating state sequence like $0, 1, 0, 1$ on a single level satisfies binary regularity and will not be penalized. Binary irregularity by inserting (e.g., $0, 1, 2, 0, 1$) or deleting (e.g., $0, 0, 1$) states are penalized.

In a hierarchical metrical transition to a level-$l$ metrical boundary, the hypermeasure positions of level $1...(l + 1)$ are updated, and the ones above level-$(l + 1)$ keep the same. Therefore, the joint transition potential is defined as
\begin{equation} \small
    \phi(z_{i-1}, z_i)=\prod_{l=1}^{L}\left\{\begin{matrix}
    A_{z_{i-1}^{(l)}z_{i}^{(l)}} & l \leq l_i + 1 \\
    \mathbb{I}[z_{i-1}^{(l)}=z_{i}^{(l)}] &  l > l_i + 1
    \end{matrix}\right.
\end{equation}
where $l_i$ denotes the corresponding metrical boundary level of $z_i$, and $\mathbb{I}[b]$ is the indicator function that returns $1$ if $b$ is true and $0$ if $b$ is false.


The emission potential function is designed as $\psi(z_i, \mathbf{p}_i)=p_{il_i}$ where $p_{il_i}$ is the $l_i$-th entry of $\mathbf{p}_i$. We use Viterbi decoding to decode the optimal hidden states $z_{1..d}$ given the observations $\mathbf{p}_{1...N}$:
\begin{equation} \small
    \mathbf{\hat{z}} = \arg \max_\mathbf{z} \psi(z_1, \mathbf{p}_1)\prod_{i=2}^N \phi(z_{i - 1}, z_{i})\psi(z_i, \mathbf{p}_i)
\end{equation}

\section{Experiments}

\subsection{Datasets}

The main dataset we use in model training and evaluation is the RWC-POP dataset. It contains 100 songs with aligned MIDI files. The MIDI files have beat and downbeat annotations that are mostly correct\footnote{2 out of 70 songs have minor beat/downbeat annotation issues.}. We manually annotated the 4-layer song-level metrical structure for 70 songs. We use 50 songs for training, 10 for validation, and 10 for testing. We acknowledge the ambiguity in data annotation which potentially causes bias by annotator subjectivity \cite{personalization}, so we publicized the annotation data and methods to facilitate discussion and future research.

All samples are quantized into 16th-note units. The binary piano roll and onset roll are extracted for each track. During training, a random 512-unit (32 measures) segment is chosen from each song, resulting in a $512\times 256$ feature matrix (128 MIDI pitches for piano roll and 128 MIDI pitches for onset roll) for each track.
To prevent overfitting, we perform label-preserving data augmentation including random pitch shift augmentation of -12 to +12 semitones and a microtiming shift up to an 8th note on the training set. The drum track does not use pitch shift augmentation and does not share the same parameter with pitched instruments for the first convolutional layer.

\subsection{Model Training}

For model training, we use a mini-batch of 16. We use the Adam optimizer \cite{kingma2014adam} on the cross-entropy classification loss with a learning rate 0.0001 for 100 epochs. In one epoch, we go through each augmented version of each song for 5 times. For each song, we randomly select 2 tracks and try to predict the song-level metrical structure given the weighted average of their predictions. We also apply dropout with a probability 0.5 after each convolutional layer to further suppress overfitting.

\subsection{Baseline Models}

While there are some existing rule-based hierarchical metrical structure analyzers \cite{tojo2013computational, jones1988rule, hiratadeepgttm} in previous works, they mostly focus on low-level (e.g., beat \& downbeat) boundary features like note transitions, durations and local rhythmic patterns. Those features are not effective enough for metrical structures above the measure level. We here build our baseline model using the methodology from \cite{Serr2012UnsupervisedDO}. For each metrical level, we calculate the similarity matrix of piano rolls on different granularity and estimate their novelty score as observations. We use a CRF decoder as mentioned above with a different set of hyper-parameters tuned for the baseline model.

To assess the necessity of introducing metrical irregularity, We also introduce another hypothetical baseline model called the \textit{oracle} model. The oracle model is not allowed to predict any hypermeter changes (i.e., it always assumes binary regularity) but it always performs the best possible prediction (i.e., maximal F1 score) for each level. This hypothetical model serves as an upper bound for any model that does not allow binary irregularity.
\begin{figure*}[ht!]
\centering
\includegraphics[width=\linewidth,clip, trim=0.0cm 5.0cm 0.0cm 1.4cm]{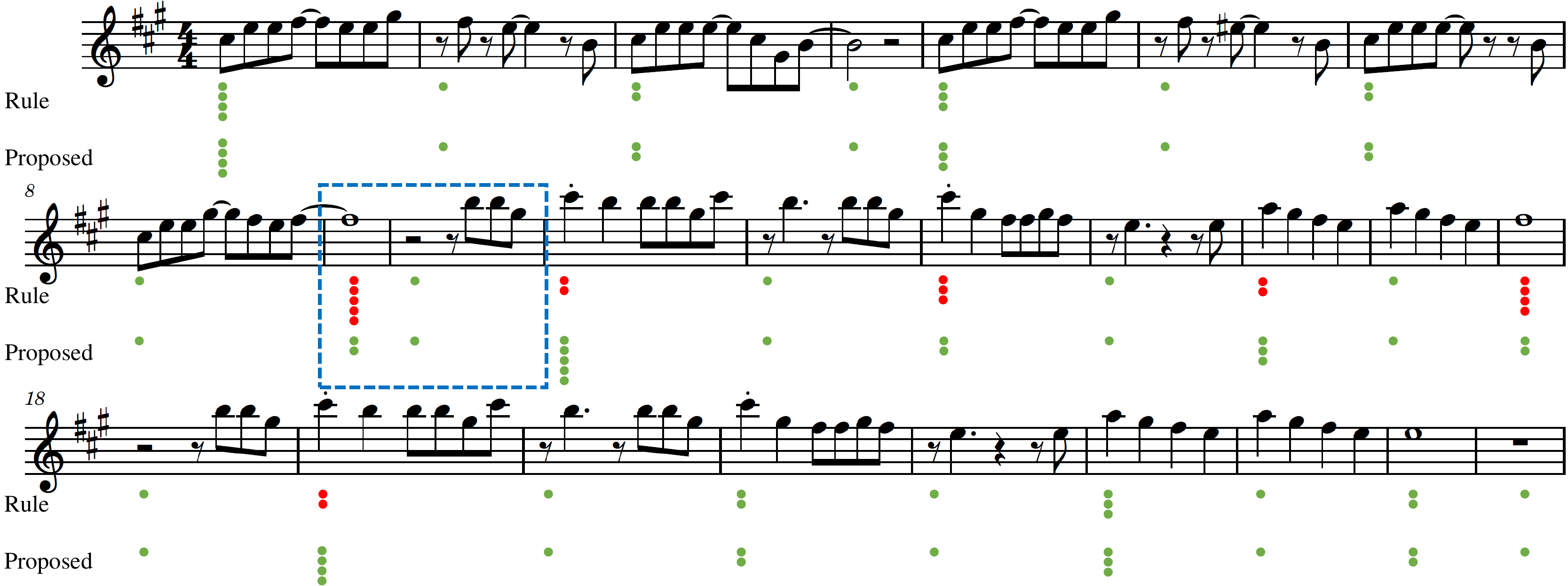}
\caption{A case study with song RWC-POP No. 008 from the test split. The song is multi-track but we only show the main melody here. The metrical structure of the song does not satisfy binary regularity because of the 2-bar extensions in the pre-chorus (marked by a dashed blue box), causing hypermeter changes. The prediction of the proposed method aligns well with the reference. The errors in the rule-based prediction are marked in red (better viewed in color).}\label{fig:case}
\end{figure*}

\subsection{Results} \label{sec:results}

Table \ref{tab:results} shows the performance of each model on the test split of RWC-POP songs. To perform a more systematic evaluation, we also created two difficult versions of each test song: (1) \textbf{no drums}: the drum track(s) are removed from the original song and each model is required to predict the same metrical structure without referring to any drum clues; (2) \textbf{mel. only}: all tracks except the melody track are removed. Each model is required to predict the structure by purely referring to the main melody track.

\begin{table}[t]
\small
\begin{tabular}{lllll}
\toprule
\textbf{Model}                                                 & \textbf{Level 1} & \textbf{Level 2} & \textbf{Level 3} & \textbf{Level 4} \\ \midrule
Proposed	& \begin{tabular}[c]{@{}r@{}}0.9848 \\ $\pm$0.0215\end{tabular}	& \begin{tabular}[c]{@{}r@{}}\textbf{0.9559} \\ $\pm$\textbf{0.0386}\end{tabular}	& \begin{tabular}[c]{@{}r@{}}\textbf{0.8880} \\ $\pm$\textbf{0.0889}\end{tabular}	& \begin{tabular}[c]{@{}r@{}}\textbf{0.6849} \\ $\pm$\textbf{0.1900}\end{tabular} \\ \midrule

\begin{tabular}[c]{@{}l@{}}Proposed\\ w/o CRF\end{tabular}	& \begin{tabular}[c]{@{}r@{}}0.9338 \\ $\pm$0.0390\end{tabular}	& \begin{tabular}[c]{@{}r@{}}0.8528 \\ $\pm$0.0937\end{tabular}	& \begin{tabular}[c]{@{}r@{}}0.7971 \\ $\pm$0.1276\end{tabular}	& \begin{tabular}[c]{@{}r@{}}0.6646 \\ $\pm$0.0844\end{tabular} \\ \midrule

Rule	& \begin{tabular}[c]{@{}r@{}}0.9228 \\ $\pm$0.0698\end{tabular}	& \begin{tabular}[c]{@{}r@{}}0.8425 \\ $\pm$0.1195\end{tabular}	& \begin{tabular}[c]{@{}r@{}}0.7485 \\ $\pm$0.1536\end{tabular}	& \begin{tabular}[c]{@{}r@{}}0.5185 \\ $\pm$0.2656\end{tabular} \\ \midrule

Oracle	& \begin{tabular}[c]{@{}r@{}}0.9427 \\ $\pm$0.1120\end{tabular}	& \begin{tabular}[c]{@{}r@{}}0.7782 \\ $\pm$0.2076\end{tabular}	& \begin{tabular}[c]{@{}r@{}}0.5188 \\ $\pm$0.1751\end{tabular}	& \begin{tabular}[c]{@{}r@{}}0.4225 \\ $\pm$0.1234\end{tabular} \\ \midrule

\begin{tabular}[c]{@{}l@{}}Proposed\\ (no drums)\end{tabular}	& \begin{tabular}[c]{@{}r@{}}\textbf{0.9868} \\ $\pm$\textbf{0.0174}\end{tabular}	& \begin{tabular}[c]{@{}r@{}}0.9519 \\ $\pm$0.0346\end{tabular}	& \begin{tabular}[c]{@{}r@{}}0.8803 \\ $\pm$0.1023\end{tabular}	& \begin{tabular}[c]{@{}r@{}}0.6611 \\ $\pm$0.2170\end{tabular} \\ \midrule

\begin{tabular}[c]{@{}l@{}}Rule\\  (no drums)\end{tabular}	& \begin{tabular}[c]{@{}r@{}}0.9312 \\ $\pm$0.0660\end{tabular}	& \begin{tabular}[c]{@{}r@{}}0.8107 \\ $\pm$0.1568\end{tabular}	& \begin{tabular}[c]{@{}r@{}}0.7055 \\ $\pm$0.2008\end{tabular}	& \begin{tabular}[c]{@{}r@{}}0.4823 \\ $\pm$0.2239\end{tabular} \\ \midrule

\begin{tabular}[c]{@{}l@{}}Proposed\\ (mel. only)\end{tabular}	& \begin{tabular}[c]{@{}r@{}}0.7413 \\ $\pm$0.2139\end{tabular}	& \begin{tabular}[c]{@{}r@{}}0.6253 \\ $\pm$0.2448\end{tabular}	& \begin{tabular}[c]{@{}r@{}}0.5551 \\ $\pm$0.2536\end{tabular}	& \begin{tabular}[c]{@{}r@{}}0.3808 \\ $\pm$0.2399\end{tabular} \\ \midrule

\begin{tabular}[c]{@{}l@{}}Rule\\  (mel. only)\end{tabular}	& \begin{tabular}[c]{@{}r@{}}0.6606 \\ $\pm$0.1451\end{tabular}	& \begin{tabular}[c]{@{}r@{}}0.4395 \\ $\pm$0.1522\end{tabular}	& \begin{tabular}[c]{@{}r@{}}0.3142 \\ $\pm$0.1211\end{tabular}	& \begin{tabular}[c]{@{}r@{}}0.1863 \\ $\pm$0.1310\end{tabular} \\ \bottomrule

\end{tabular}
\caption{Evaluated F1 scores on the test split of the RWC-POP dataset. }
\label{tab:results}
\end{table}

From the results, we can see that our proposed model performs better than the rule-based counterpart on all metrical levels. Since most test songs have more than 10 MIDI tracks, they provide sufficient metrical hints to both the proposed model and the rule-based model even if the drum track is removed. When we only have the melody track, both the proposed model and the rule-based model's performances are not satisfactory even on the first level beyond measure. Still, the data-driven approach shows improved performance compared to the rule-based system.

We observe that the proposed model is better at capturing irregular metrical structures than the rule-based approach. Figure \ref{fig:case} shows a cherry-picked example where binary irregularity can be found. Both the proposed model and the rule-based baseline can detect such irregularity but only the proposed model correctly tells the exact position of the hypermetrical change.

There is also a tendency for the performance to drop rapidly from lower to higher levels. We believe there are 2 main reasons. First, the higher levels have fewer positive samples, making it hard for the model to learn its semantic characteristics. Second, metrical structures on higher levels are often more ambiguous than lower ones even for human listeners. Sometimes, the highest level (level 4) needs to decide how to group parts together (e.g., verse + pre-chorus or pre-chorus + chorus). Different decisions are sometimes all acceptable.

\subsubsection{Out of Distribution Evaluation}

We also want to know whether the proposed model can be applied to MIDI files with very different orchestration setups. Such experiments are hard to perform because of the lack of ground truth annotations. We here perform a small-scale experiment on the POP909 \cite{ziyu_wang_2020_4245366} dataset. We select the first 5 songs in the dataset ordered by index and manually annotate the metrical structure\footnote{We are aware that POP909's downbeat annotations are sometimes inaccurate and we manually fixed them.}. POP909 is a dataset of Chinese pop songs rearranged for piano performance. Each song only has 3 tracks, i.e., a vocal track and two piano tracks, making it harder compared to RWC-POP. The results are shown in table \ref{tab:resultspop909}.

From the results, we can see the performance degrades even when all 3 tracks are present. By case inspection, we find that the proposed model has generally satisfactory performance on 3 out of 5 songs on lower layers. However, there is one complex song\footnote{POP909 No. 005: \textit{I Believe} by Van Fan.} with multiple metrical and hypermeter changes that make all the approaches fail. Also, due to the lack of rhythmic clues (e.g., drums), a deeper understanding of the syntax and semantics of melody and chords might be required to perform musically meaningful segmentation, which we assume our model can hardly acquire on a small training set of 50 pop songs.

\begin{table}[t]
\small
\begin{tabular}{lllll}
\toprule
\textbf{Model}                                                 & \textbf{Level 1} & \textbf{Level 2} & \textbf{Level 3} & \textbf{Level 4} \\ \midrule
Proposed	& \begin{tabular}[c]{@{}r@{}}\textbf{0.9084} \\ $\pm$\textbf{0.0896}\end{tabular}	& \begin{tabular}[c]{@{}r@{}}\textbf{0.8742} \\ $\pm$\textbf{0.1101}\end{tabular}	& \begin{tabular}[c]{@{}r@{}}\textbf{0.6470} \\ $\pm$\textbf{0.2174}\end{tabular}	& \begin{tabular}[c]{@{}r@{}}0.4930 \\ $\pm$0.3132\end{tabular} \\ \midrule

Rule	& \begin{tabular}[c]{@{}r@{}}0.6818 \\ $\pm$0.1855\end{tabular}	& \begin{tabular}[c]{@{}r@{}}0.6625 \\ $\pm$0.1550\end{tabular}	& \begin{tabular}[c]{@{}r@{}}0.5163 \\ $\pm$0.1195\end{tabular}	& \begin{tabular}[c]{@{}r@{}}0.3446 \\ $\pm$0.1856\end{tabular} \\ \midrule

Oracle	& \begin{tabular}[c]{@{}r@{}}0.8527 \\ $\pm$0.1735\end{tabular}	& \begin{tabular}[c]{@{}r@{}}0.7348 \\ $\pm$0.2763\end{tabular}	& \begin{tabular}[c]{@{}r@{}}0.5883 \\ $\pm$0.2493\end{tabular}	& \begin{tabular}[c]{@{}r@{}}\textbf{0.5767} \\ $\pm$\textbf{0.2474}\end{tabular} \\ \midrule
   
\begin{tabular}[c]{@{}l@{}}Proposed\\ (mel. only)\end{tabular}	& \begin{tabular}[c]{@{}r@{}}0.6742 \\ $\pm$0.2962\end{tabular}	& \begin{tabular}[c]{@{}r@{}}0.6542 \\ $\pm$0.2737\end{tabular}	& \begin{tabular}[c]{@{}r@{}}0.5685 \\ $\pm$0.2403\end{tabular}	& \begin{tabular}[c]{@{}r@{}}0.4797 \\ $\pm$0.2053\end{tabular} \\ \midrule

\begin{tabular}[c]{@{}l@{}}Rule\\ (mel. only)\end{tabular}	& \begin{tabular}[c]{@{}r@{}}0.6062 \\ $\pm$0.1034\end{tabular}	& \begin{tabular}[c]{@{}r@{}}0.3933 \\ $\pm$0.0275\end{tabular}	& \begin{tabular}[c]{@{}r@{}}0.2642 \\ $\pm$0.0546\end{tabular}	& \begin{tabular}[c]{@{}r@{}}0.1551 \\ $\pm$0.0305\end{tabular} \\ \bottomrule

\end{tabular}
\caption{Evaluated F1 scores on the first 5 songs in the POP909 dataset. Mel. only denotes that the melody track is used. Otherwise, all 3 tracks (melody, bridge and piano) are used.}
\label{tab:resultspop909}
\end{table} 

\subsection{Confidence Score Analysis}


To perform a statistical analysis of the model behavior and provide a musicological view of the model prediction, we perform model inferences on a large selection of the Lakh MIDI dataset \cite{raffel2016learning}. To ensure enough accuracy of model prediction, we only select a part of the Lakh MIDI dataset that has a similar orchestration compared to RWC-POP. We filter the MIDI files according to the following criteria: (1) it contains at least 6 MIDI tracks, including 1 drum track and 1 track whose name contains strings "melody" or "vocal"; (2) if multiple MIDI files' identified audio sources are the same, at most one MIDI file is kept. A filtered dataset of 3,739 MIDI files is collected. 

We here evaluate the relevance of instruments and the model's predicted confidence score. Notice that the instrument program number is not a part of the model input, so the only difference comes from the rhythmic properties of their scores. We collect the unnormalized confidence scores $\alpha^{(t)}$ for each track $\mathbf{m}_t$ of different instruments, and calculate their means and standard derivations. Specially, we regard all melody tracks (identified by their names) as a new instrument and ignore its original MIDI program number. Also, we remove all tracks with too many measure-level rests (more than 1/3 of the whole song) since they trivially result in low confidence scores.

Table \ref{tab:ins} shows the results of confidence score analysis. We can see that drums are the strongest clue for metrical structures. The melody track and many melodic instruments (e.g., guitars) also serve as useful clues. On the other hand, instruments that produce slow accompaniments (e.g., string ensemble and pads) are less preferred.

\begin{table}[t]
\small
\centering
\begin{tabular}{ll}
\toprule
\textbf{Track/Instrument} & \textbf{Confidence}   \\ \midrule
Melody               & 1.78 $\pm$ 1.22 \\ \midrule
Drum                 & 3.61 $\pm$ 2.58 \\ \midrule
Acoustic Grand Piano & 0.35 $\pm$ 1.69 \\ \midrule
Electric Guitar (jazz) &	0.75 $\pm$ 1.55 \\ \midrule
Acoustic Bass        & 0.33 $\pm$ 1.66 \\ \midrule
String Ensemble    & 0.11$ \pm$ 1.70 \\ \midrule
Pad (warm)	& -0.86 $\pm$ 1.78 \\ \bottomrule
\end{tabular}
\caption{A selected view of the means and standard derivations of the confidence score for different tracks/instruments. The melody track is identified by its name instead of the MIDI instrument. The drum track is identified by its MIDI channel number (No. 10).}
 \label{tab:ins}
\end{table}

\subsubsection{Drum Track Analysis}

As another experiment of musicological analysis, we perform an experiment on the relation between drum notes and the metrical structure level. For simplicity, we only collect samples that a certain drum event happens exactly on the downbeat, and we collect the corresponding metrical boundary level of that downbeat. The results are shown in Table \ref{tab:drum}. While the occurrence of many drum events does not significantly change the distribution of the metrical boundary level, the crash cymbal and splash cymbal are certainly useful clues to a high-level metrical boundary. This aligns with people's perception of them since these cymbals are usually associated with a strong burst of energy, serving as an important rhythmic hint.

\begin{table}[t]
\small
\centering
\begin{tabular}{llllll}
\toprule
\textbf{Drums (\%)} & \textbf{L-0} & \textbf{L-1} & \textbf{L-2} & \textbf{L-3} & \textbf{L-4} \\ \midrule
Any            & 50.00 &	24.71 &	12.06 &	6.21 &	7.02
           \\ \midrule
Bass Drum      & 48.66 &	25.10 &	12.46 &	6.47 &	7.30
           \\ \midrule
Acoustic Snare & 52.27 &	23.28 &	11.19 &	6.27 &	7.00
           \\ \midrule
Closed Hi Hat  & 51.32 &	25.14 &	11.91 &	5.54 &	6.11
           \\ \midrule
Open Hi Hat    & 51.90 &	25.07 &	11.33 &	5.38 &	6.32
           \\ \midrule
Crash Cymbal   & 20.97 &	19.13 &	18.58 &	18.94 &	22.38
          \\ \midrule
Ride Cymbal    & 51.41 &	25.03 &	11.57 &	5.61 &	6.39
          \\ \midrule
Splash Cymbal    & 34.80 &	22.30 &	16.00 &	12.90 &	14.01
         \\ \bottomrule

\end{tabular}
\caption{A selected view of the frequency of different drum instruments (on a downbeat) associated with different levels of metrical boundaries. L-$n$ means level-$n$ metrical boundary.}
\label{tab:drum}
\vspace{-7pt}
\end{table}

\section{Conclusion and Future Work}

In this paper, we propose a data-driven approach for hierarchical metrical structure analysis of symbolic music. Our model adopts a TCN-CRF architecture and accepts an arbitrary number of voices as input.  Experiments on MIDI datasets show that our model performs better than rule-based methods under different orchestration settings.

The model performance is still not satisfactory, especially for high-level metrical structures and music with very different orchestration. We assume the performance would be better if more data were annotated, but there are also other possible directions for data-driven methods. First, self-supervised or semi-supervised methods might be a helpful complement to the lack of labeled datasets. For example, a consistency loss can be used to evaluate the prediction between different voices in the same music piece. Different data augmentation strategies might also be helpful. Second, it might be useful to utilize datasets of related tasks (e.g., section labels) as a source of weak supervision. Related tasks can also be used for multi-task learning.

Other potential future works include improving the automatic analysis system of other hierarchical structures, e.g., the grouping structures. The application of hierarchical structure analysis in the audio domain is also worth exploring.

\bibliography{ISMIRtemplate}

%
%
%
%
%

\end{document}